\documentclass[aps,prl,twocolumn,superscriptaddress,showpacs,amsmath,amssymb]{revtex4-1}
\usepackage[utf8]{inputenc}
\usepackage{amsmath,amsthm}
\usepackage{graphicx}
\usepackage[colorlinks = true,
            linkcolor = blue,
            urlcolor  = blue,
            citecolor = blue,
            anchorcolor = blue]{hyperref}
\usepackage{bbold}
\usepackage{braket}
\usepackage{xcolor}
\usepackage[title]{appendix}
\newtheorem{conj}{Conjecture}
\newtheorem{prop}{Proposition}
\newtheorem{lema}{Lemma}

\newcommand{\U}{\ensuremath{\mathcal{U}}}
\newcommand{\J}{\ensuremath{\mathcal{J}}}
\newcommand{\x}{{x}}
\newcommand{\y}{{y}}
\newcommand{\z}{{z}}
\renewcommand{\S}{\mathbf{S}}
\renewcommand{\L}{\mathbf{L}}
\newcommand{\T}{\mathbf{T}}
\newcommand{\V}{\mathcal{V}}
\newcommand{\lhalf}{{\textstyle\frac{\lambda}{2}}}
\newcommand{\mhalf}{{\textstyle\frac{\mu}{2}}}
\newcommand{\tr}[1]{\ensuremath{\operatorname{tr}\!{#1}}}
\renewcommand{\d}[1]{\ensuremath{\operatorname{d}\!{#1}}}
\newcommand{\real}{{\rm Re}}
\newcommand{\imag}{{\rm Im}}
\newcommand{\Gd}{{\rm Gd}}

\begin{document}

    \title{Ballistic spin transport in a periodically driven integrable quantum system}
    \author{Marko Ljubotina}
    \affiliation{Faculty of Mathematics and Physics, University of Ljubljana, Slovenia}
    \author{Lenart Zadnik}
    \affiliation{Faculty of Mathematics and Physics, University of Ljubljana, Slovenia}
    \author{Toma\v z Prosen}
    \affiliation{Faculty of Mathematics and Physics, University of Ljubljana, Slovenia}
    
    \begin{abstract}
        We demonstrate ballistic spin transport of an integrable unitary quantum circuit, which can be understood either as a paradigm of an integrable periodically driven (Floquet) spin chain, or as a Trotterized anisotropic ($XXZ$) Heisenberg spin-1/2 model. We construct an analytic family of quasi-local conservation laws that break the spin-reversal symmetry and compute a lower bound on the spin Drude weight which is found to be a fractal function of the anisotropy parameter. Extensive numerical simulations of spin transport suggest that this fractal lower bound is in fact tight.
    \end{abstract}
    \maketitle
    
    {\bf Introduction.--}
    Understanding transport in various out-of-equilibrium setups, in particular in low dimensions, is one of the main challenges of theoretical condensed matter physics \cite{affleck}. Experimental evidence corroborates the seemingly controversial proposal~\cite{castella,zotos} that integrable systems generically exhibit ballistic transport even at high temperatures~\cite{experimental1,experimental2,FHM1}. This proposal has received rigorous justification in terms of the existence of an extensive number of quasi-local conserved quantities \cite{ip13,tomaz11,tomaz13,tomaz14,pereira14,review} which form a basis for the hydrodynamic theory of interacting integrable systems \cite{bertini16,doyon16,deluca17}.
    
    Recently, periodically driven (Floquet) spin chains with local interactions have attracted considerable attention. This was in particular due to the possibility of exhibiting generalized thermalization towards non-equilibrium steady states \cite{ggefloquet} and distinct dynamical phases with respect to the spontaneous breaking of time-translation invariance \cite{nayak,sondhi}. Still, the possibility of strictly ballistic transport in interacting quantum integrable Floquet systems has never been explored (see~\cite{katja,sarang} for a classical lattice setting), even though a peculiar robustness of transport to integrability breaking has been observed a while ago \cite{tp98,tp98ql} (also~\cite{arnab}).
    
    For concreteness, let us consider spin transport. Without resorting to the spectroscopic approach~\cite{FHM2}, which is harder to justify in Floquet systems,
    ballistic transport can be defined as a linear growth of the spin current in time, after the system has been prepared in an initial state supporting a small gradient of magnetization. This can be formulated in terms of a nonzero {\em Drude weight}
    \begin{align}
        D=\lim_{t\to \infty}\lim_{N\to \infty} \lim_{\mu\to 0}\frac{\langle J(t)\rangle_\mu}{2Nt\mu},
        \label{eq:drude_weight}
    \end{align}
    where $J=\sum_n j_n$ is the extensive spin current operator on a spin chain of length $N$ and $J(t)$ its time dependence. $\langle\bullet\rangle_\mu$ denotes the average in the initial state with a small gradient of magnetization $\mu$, say $\rho_\mu \sim \exp\left(\mu\sum_n n\,\sigma^z_n\right)$, where $\sigma^z_n$ is a local spin variable. A formula similar to~\eqref{eq:drude_weight} holds even if the system is initially prepared in two equilibrated halves at different magnetizations $\mu_L$ and $\mu_R$ with $\mu\sim(\mu_L-\mu_R)/N$ representing the effective gradient \cite{vasseur15,karrasch17,ilievski17}. This partitioned initial state is easier to simulate using state-of-the-art tensor network simulations.
    
    Expanding to the first order in $\mu$, the Drude weight can be expressed solely in terms of equilibrium auto-correlation functions using the {\em Kubo formula}, see Appendix A of the Supplemental material (SM)~\cite{supp}. This can in turn be bounded from below by means of the {\em Mazur inequality} \cite{M69,suzuki,zotos} (see \cite{ip13} for a rigorous derivation 
    in extended systems)
    \begin{align}
    \begin{aligned}
        D=&\lim_{t\to \infty}\lim_{N\to \infty}\frac{1}{2N}\frac{1}{t}\sum_{\tau=1}^t\,\langle J J(\tau)\rangle\\
        &\ge\lim_{N\to\infty}\frac{1}{2N}\sum_k \frac{|\langle J,Q_k \rangle|^2}{\langle Q_k,Q_k \rangle}.
        \label{eq:kubo}
    \end{aligned}
    \end{align}
    Here $Q_k$ are conserved quantities orthogonal with respect to the inner product $\langle A,B\rangle=\tr[A^\dagger B]/2^N$, assuming that the reference equilibrium state is the
    maximum entropy state $\rho_{\mu=0}=2^{-N}\mathbb{1}$ and the local Hilbert space dimension is $2$. In order for the bound to be finite the conserved quantities should be linearly extensive or {\em quasi-local}, $\langle Q_k,Q_k\rangle\propto N$, and should have a finite overlap with the spin current, $\langle J,Q_k\rangle\ne 0$. For the latter to hold, $Q_k$ must {\em not} be symmetric, $\mathcal{P}Q_k\mathcal{P}\neq Q_k$, with respect to the spin-reversal operator $\mathcal{P}$, which flips the spin current ${\cal P}J{\cal P} = -J$. In the easy-plane (gapless) regime of the Heisenberg $XXZ$ model, for example, the integrals of motion with all of the required properties have been shown to exist~\cite{tomaz11,tomaz13,tomaz14,pereira14}. 
    
    In the present paper, we aim to rigorously establish a regime of ballistic transport in a Floquet driven integrable model related to the Trotterized $XXZ$ spin-$1/2$ chain. We introduce the dynamical protocol as a local quantum circuit, establish its connection with the six-vertex $R$-{\em matrix} and integrability structure of the $XXZ$ model, and define the spin currents and continuity equations arising from the global $U(1)$ symmetry of the model. Despite its driven nature, we construct a set of quasi-local conservation laws which break the spin-reversal symmetry. We then show how to evaluate the optimized Mazur lower bound on the spin Drude weight. Extensive numerical simulations using the time-evolving block decimation (TEBD) algorithm  strongly suggest that this  bound, which is a fractal function of parameters, is in fact saturated, similarly as in the continuous-time case \cite{ilievski17,klumper19}.

    {\bf The model.--} Consider a spin-1/2 chain with $N\in2\mathbb{Z}$ sites and periodic boundary conditions. The local {\em physical} space on each site will be denoted by $\V_p\equiv\mathbb{C}^2$. We are interested in a discrete-time Liouville-von Neumann equation for a density matrix $\rho_{t+1}=\U\,\rho_t\,\U^\dagger$.
    \begin{figure}[ht!]
        \centering
        \hspace{-0.25cm}
        \includegraphics[width=0.85\linewidth]{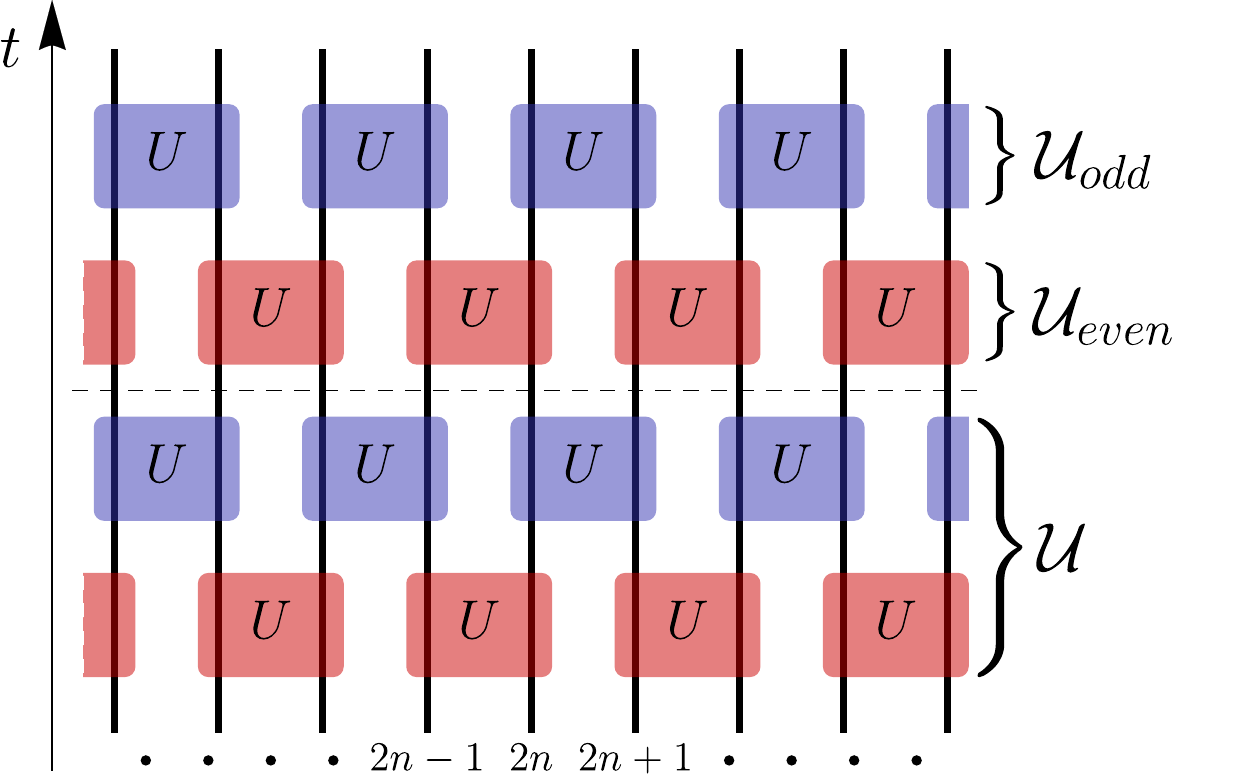}
        \caption{Schematics of the time evolution. The red gates represent $\U_{even}$ and the blue ones $\U_{odd}$. The direction of the time is upwards. The schematic shows two full time steps in the bulk of the system.}
        \label{fig:prop_scheme}
    \end{figure}
    The propagator $\U=\U_{odd}\,\U_{even}$ acts in two steps
    \begin{align}
    \begin{aligned}
        \U_{odd}=\prod_{n=1}^{N/2}U_{2n,2n+1}\,,\qquad 
        \U_{even}=\prod_{n=1}^{N/2}U_{2n-1,2n}\,,
        \label{eq:propagation}
    \end{aligned}
    \end{align}
    where 
    \begin{align}
        U_{n,n+1}=e^{-i\,\J\,(\sigma^\x_n\sigma^\x_{n+1}+\sigma^\y_n\sigma^\y_{n+1})-i\,\J'\,(\sigma^\z_n\sigma^\z_{n+1}-\mathbb{1})}
        \label{eq:propagator}
    \end{align} 
    is a unitary gate acting on two neighbouring sites labelled with $n$ and $n+1$, see Figure~\ref{fig:prop_scheme}. Here $\sigma^\alpha$ ($\alpha=\x,\y,\z$) are Pauli matrices. By considering infinitesimally small couplings $\J$ and $\J'$ and an infinite number of time steps we recover the continuous-time dynamics of the $XXZ$ model, according to the Trotter-Suzuki formula. 
    
    The local 2-site unitary gate can be rewritten as $U_{n,n+1}=\check{R}_{n,n+1}(\lambda)$ where
    \begin{align}
        \check{R}(\lambda)=
        \left(\,
        \begin{matrix}
        1 & 0 & 0 & 0 \\
        0 & \frac{\sin\eta}{\sin(\lambda+\eta)} & \frac{\sin\lambda}{\sin(\lambda+\eta)} & 0 \\
        0 & \frac{\sin\lambda}{\sin(\lambda+\eta)} & \frac{\sin\eta}{\sin(\lambda+\eta)} & 0 \\
        0 & 0 & 0 & 1 \\
        \end{matrix}\,
        \right)
        \label{eq:rmatrix}
    \end{align}
    denotes the braid form of the $R$-matrix of the $XXZ$ model. The new parameters $\eta$ and $\lambda$ can be implicitly expressed as unique functions of $\J$ and $\J'$ through the following pair of relations
    \begin{align}
        e^{2 i(\J\pm \J')}=\frac{\sin\eta-\sin\lambda}{\sin(\eta\pm\lambda)}.
        \label{eq:coordinates}
    \end{align}
    
\begin{figure}[ht!]
        \centering
        \hspace{-0.65cm}
        \includegraphics[width=0.75\linewidth]{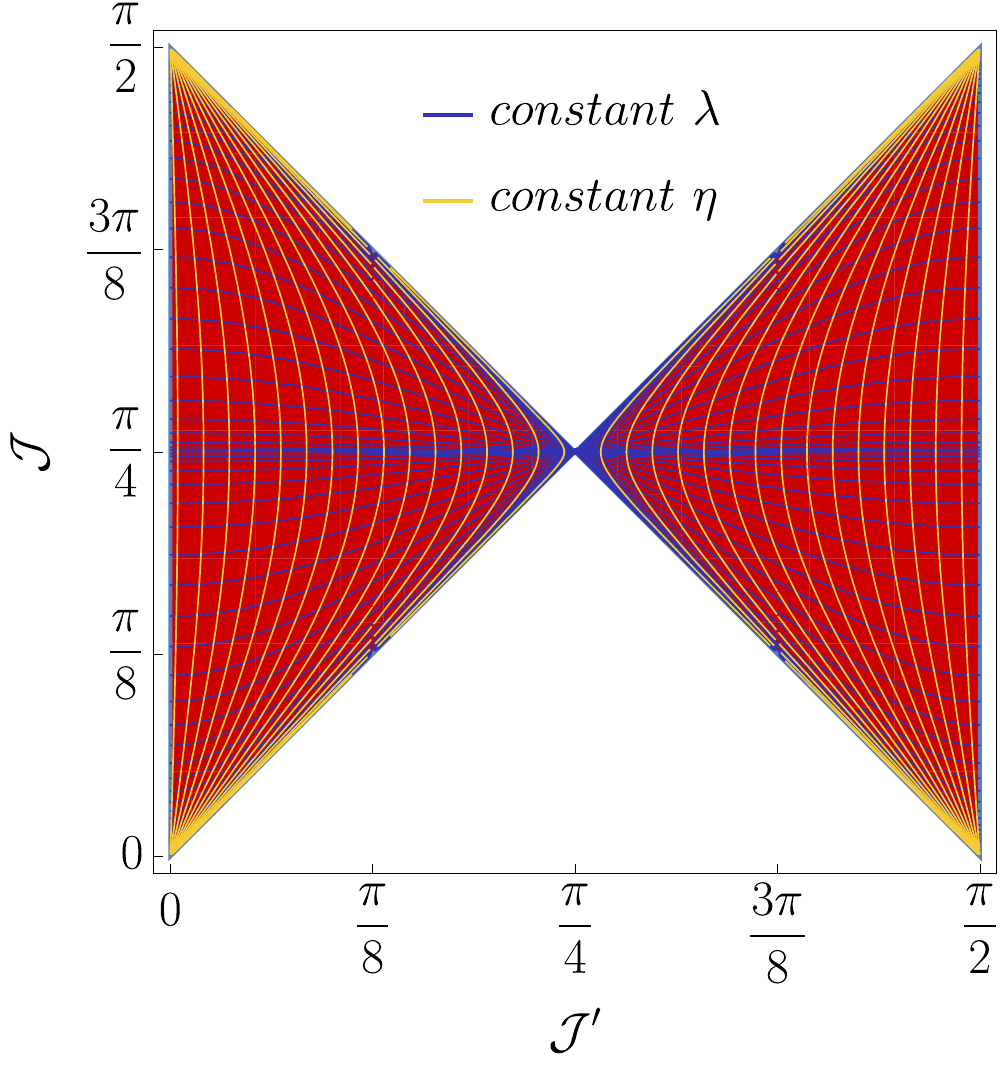}
        \caption{The coloured area corresponds to real $\eta$ and imaginary $\lambda$. The blue and yellow lines are constant $\lambda$ and constant $\eta$ contours, respectively. The continuous time limit corresponds to $\J,\,\J'\to 0$. Note: $\eta=\pi/2$ corresponds to the free model ($\J'=0$). In the limit $\lambda\to\infty$ ($\J\to\pi/4$) the local propagator (\ref{eq:propagator}) reduces to a SWAP gate with some $\J'$-dependent phase.} 
        \label{fig:coord}
    \end{figure}

    The continuous-time limit is recovered as an expansion in small $\lambda$ which gives $U_{n,n+1}=\mathbb{1}+\lambda\,h_{n,n+1}+\mathcal{O}(\lambda^2)$ with the local Hamiltonian density
    \begin{align}
        h_{n,n+1}=\frac{1}{2 \sin\eta}\big(\sigma_n^\x\sigma_{n+1}^\x+\sigma_n^\y\sigma_{n+1}^\y+\Delta\,(\sigma_n^\z\sigma_{n+1}^\z-\mathbb{1})\big),
        \label{eq:hamiltonian}
    \end{align}
    with $\Delta=\cos\eta$ being the anisotropy parameter. Clearly, real $\eta$ and imaginary $\lambda$ correspond to the gapless/easy-plane regime, shown in Figure~\ref{fig:coord}, whereas imaginary $\eta$ and real $\lambda$ correspond to the gapped/easy-axis regime. 
    
\begin{figure}[ht!]
        \centering
        \includegraphics[width=1.0\linewidth]{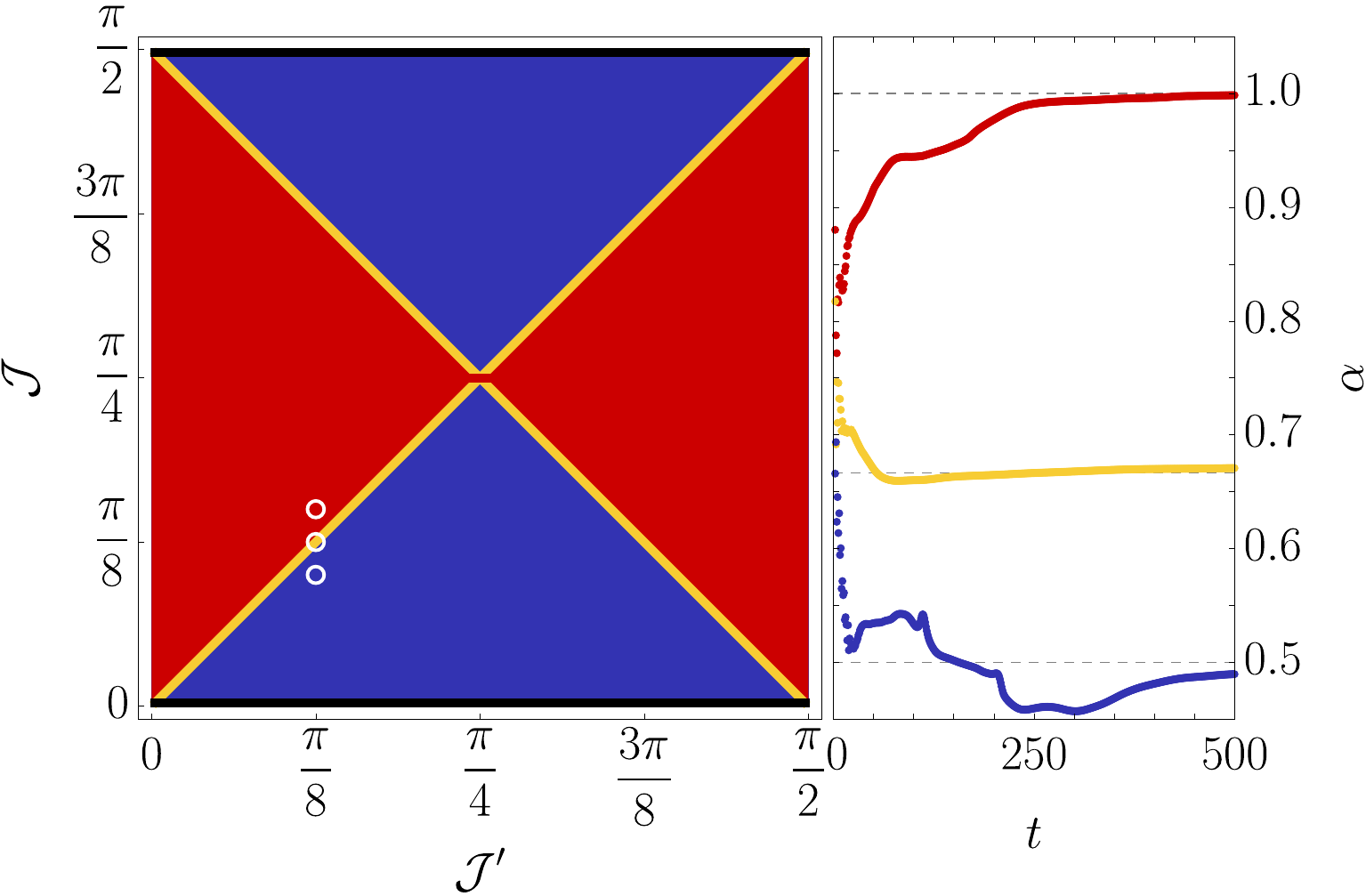}
        \caption{A schematic phase diagram of the model based on TEBD simulations. The three circles mark the values of $\J$ and $\J'$ used in the right plot, which in turn depicts the time-dependence of the exponent $\alpha$, defined through the transport of magnetization between two half-chains \cite{ljubotina} $\alpha(t)=({\rm d}/{\rm d}\log t)\log\left( \sum_{\tau=0}^t \langle j_{N/2+1}(\tau)\rangle\right)$. We can recognize ballistic, super-diffusive and diffusive regimes (red, yellow and blue respectively). Here we have used bond dimension $64$ on a chain of length $N=3600$.}
        \label{fig:phase}
    \end{figure}

    In the gapless regime of the continuous-time limit ($|\Delta|<1$) the Drude weight has rigorously been shown to be nonzero for a dense set of anisotropies parametrized by $\eta=l\,\pi/m$, where $l$ and $m$ are coprime integers~\cite{tomaz13,tomaz14}. In this paper we extend this discussion for the same set of anisotropies to a discrete time, i.e., to all imaginary $\lambda$. This will cover the ballistic regime in the phase diagram of our model, shown in red in Figure~\ref{fig:phase}, which is determined by adapting the numerical method of Ref.~\cite{ljubotina}. We stress that we observe ballistic transport for any ratio ${\cal J}'/{\cal J}$, even for $|\J'|>|\J|$, unlike in the continuous-time case. 
    Note also that the other two transport regimes can be clearly established numerically -- the super-diffusive in yellow and the diffusive in blue.
    
    {\bf Spin currents.--} Due to the $U(1)$ symmetry of the propagator~\eqref{eq:rmatrix} the total magnetization $M=\sum_{n=1}^N\sigma^\z_n$ is a conserved quantity. As a result of the discrete time propagation we identify two continuity equations, separately for odd/even sites
    \begin{align}
    \begin{aligned}
        &\U^\dagger\,\sigma_{2n+1}^\z\,\U-\sigma_{2n+1}^\z=-j^{}_{2n+2}+j'_{2n+1},\\
        &\U^\dagger\,\sigma_{2n}^\z\,\U-\sigma_{2n}^\z=-j'_{2n+1}+j^{}_{2n}
        \label{eq:scont}
    \end{aligned}
    \end{align}    
    Through them we can define two local current densities, the {\em even} current $j^{}_{2n}$ and the {\em odd}, $j'_{2n+1}$. The even one is
    \begin{align}
    \begin{gathered}
        j^{}_{2n}=\frac{4\sin\lambda\sin\eta}{\cos2\eta-\cos2\lambda}\,\big(\sigma^+_{2n-1}\,\sigma^-_{2n}-\sigma^-_{2n-1}\,\sigma^+_{2n}\big)+\\
        +\frac{2\,(\sin\lambda)^2}{\cos2\eta-\cos2\lambda}\,\big(\sigma^\z_{2n-1}-\sigma^\z_{2n}\big)
        \label{eq:localJ}
    \end{gathered}
    \end{align}
    while the odd current can be computed as $j_{2n+1}'=\,\U_{even}^\dagger\,j^{}_{2n+1}\,\U_{even}$ and operates on four adjacent sites. In the continuous-time limit both local currents reduce to
    \begin{align}
        j^{}_{n}=j'_{n}=-\frac{2\lambda}{\sin\eta}\big(\sigma^+_{n-1}\,\sigma^-_{n}-\sigma^-_{n-1}\,\sigma^+_{n}\big)+\mathcal{O}(\lambda^2),
        \label{eq:cont_current}
    \end{align}
    with the pre-factor $(\sin\eta)^{-1}$ coming from the Hamiltonian~\eqref{eq:hamiltonian}.
    
    The total extensive spin current is now defined as $J=\sum_{n=1}^{N/2}(j^{}_{2n}+j'_{2n+1})$. It is clearly anti-symmetric under spin-reversal $\mathcal{P}=\prod_{n=1}^N\sigma_n^\x$, i.e., $\mathcal{P}J\mathcal{P}=-J$. We now proceed to construct the relevant conservation laws for all imaginary $\lambda$ and for a dense set of commensurate values of the anisotropy parameter $\eta=l\,\pi/m$.
    
    {\bf Quasilocal integrals of motion.--} The construction of anti-symmetric conservation laws is similar as in the continuous-time limit~\cite{tomaz14}. However, they are now generated by the staggered transfer operator
    \begin{align}
      T(\varphi,s)=\tr{_a}\Big[\prod_{n=1}^N \L_{n,a}\big(\varphi-(-1)^n\,\lhalf,s\big)\Big].
      \label{eq:transfer}
    \end{align}
    Here, $\L_{n,a}(\varphi,s)$ denotes the Lax operator acting on the $n$-th physical space $\V_p$ in the string $\bigotimes_{n=1}^N \V_p$ as a $2\times 2$ matrix
    \begin{align}
        \L(\varphi,s)=\frac{1}{\sin\varphi}
        \begin{pmatrix}
        \sin(\varphi+\eta\,\S_s^\z) & \sin(\eta)\,\S_s^- \\
        \sin(\eta)\,\S_s^+ & \sin(\varphi-\eta\,\S_s^\z)
        \end{pmatrix}
        \label{eq:lax}
    \end{align}
    whose elements are themselves matrices in the {\em auxiliary space} $\V_a$.
    For $\eta=l\,\pi/m$ the latter is an $m$-dimensional complex spin-$s$ representation of the quantum group $\mathcal{U}_q(sl_2)$ ($q=e^{i\eta}$) traced out in the final expression~\eqref{eq:transfer}. Its generators have an explicit form reminiscent of the angular momentum generators
    \begin{align}
    \begin{aligned}
        &\S_s^\z=\sum_{k=0}^{m-1}(s-k)\ket{k}\bra{k},\\
        &\S_s^+=\sum_{k=0}^{m-2}\frac{\sin(k+1)\eta}{\sin\eta}\ket{k}\bra{k+1},\\
        &\S_s^-=\sum_{k=0}^{m-2}\frac{\sin(2s-k)\eta}{\sin\eta}\ket{k+1}\bra{k}.
        \label{eq:Uqsl2}
    \end{aligned}
    \end{align}
    Together with $\check{R}(\lambda)$ given in~\eqref{eq:rmatrix}, the Lax operator~\eqref{eq:lax} satisfies the Yang-Baxter equation which implies $[T(\varphi,s),\U]=0$ and $[T(\varphi,s),T(\varphi',s)]=0$ (see Appendices B and C of the SM~\cite{supp}). 
    
    For $\lambda=0$ the spin-reversal asymmetric conservation laws of the $XXZ$ spin $1/2$ chain were 
    previously produced~\cite{tomaz13,tomaz14,pereira14} as
   \begin{equation}
     Z(\varphi)=\frac{1}{2\eta\sin\eta}\,\partial_s T(\varphi,s)\mid_{s=0},
    \end{equation}
    and shown to be linearly extensive ({\em quasi-local}) inside an analyticity strip $|\real\,\varphi-\frac{\pi}{2}|<\frac{\pi}{2m}$. Here we will simply show that this expression can be extended to arbitrary values of
    parameter $\lambda$ if the staggered form \eqref{eq:transfer} of the transfer matrix is used. Since $\lambda$ is purely imaginary, the region of quasi-locality remains the same. The detailed construction of these conservation laws for finite $\lambda$ is presented in Appendix D of the SM~\cite{supp}. 
    
    In order to maximize the Mazur lower bound~\eqref{eq:kubo} for the spin Drude weight we need to minimize the norm of the conservation laws $Q_k\sim Z(\varphi)$ without reducing the overlap with the spin current $\langle J,Q_k\rangle$.
    Due to the asymmetry of the current operator $\mathcal{P}J\mathcal{P}=-J$, only the spin-reversal anti-symmetric component $Z^-(\varphi)=\frac{1}{2}(Z(\varphi)-\mathcal{P}Z(\varphi)\mathcal{P})$ contributes to the lower bound. Furthermore, we can subtract a term proportional to the total magnetization,
    $Z^{-}_\perp(\varphi)=Z^{-}(\varphi)-\frac{1}{N}\langle M,Z(\varphi)\rangle\,M$,
    since the latter is orthogonal to the spin current, see Appendix E of the SM~\cite{supp}. The overlap between $Z^-_\perp(\varphi)$ and the current is now given by
    \begin{align}
        j(\varphi)=\lim_{N\to\infty}\frac{1}{N}\langle Z(\bar{\varphi}),J\rangle=\frac{\sin\lambda}{(\cos\lambda-\cos2\varphi)\,\sin\eta}.
        \label{eq:overlap}
    \end{align}
    The quasi-locality of $Z^-_\perp(\varphi)$ is seen from the $N$-independence of $K(\varphi,\varphi')\equiv\lim_{N\to\infty}\frac{1}{N}\langle Z_\perp^-(\bar{\varphi}),Z_\perp^-(\varphi')\rangle$ with the following conjectured analytical result
    \begin{widetext}
    \begin{align}
    \begin{gathered}
        K(\varphi,\varphi')=\frac{\big(\cos(\varphi-\varphi'+\lambda)+\cos(\varphi-\varphi'-\lambda)-2\cos(\varphi+\varphi')\big)\,\sin[(m-1)(\varphi+\varphi')]+(\sin\lambda)^2\,\sin[m(\varphi+\varphi')]}{4\,(\sin\eta)^2\,(\cos2\varphi-\cos\lambda)\,(\cos\lambda-\cos2\varphi')\,\sin[m(\varphi+\varphi')]}.\label{eq:norms}
    \end{gathered}
    \end{align}
    \end{widetext}
    Expressions $j(\varphi)$ and $K(\varphi,\varphi')$, given by \eqref{eq:overlap} and \eqref{eq:norms} respectively, are the essential ingredients for the lower bound on the Drude weight, which we discuss next.
    
\begin{figure}[ht!]
        \centering
        \includegraphics[trim={0.1cm 0 0 0},clip,width=1\linewidth]{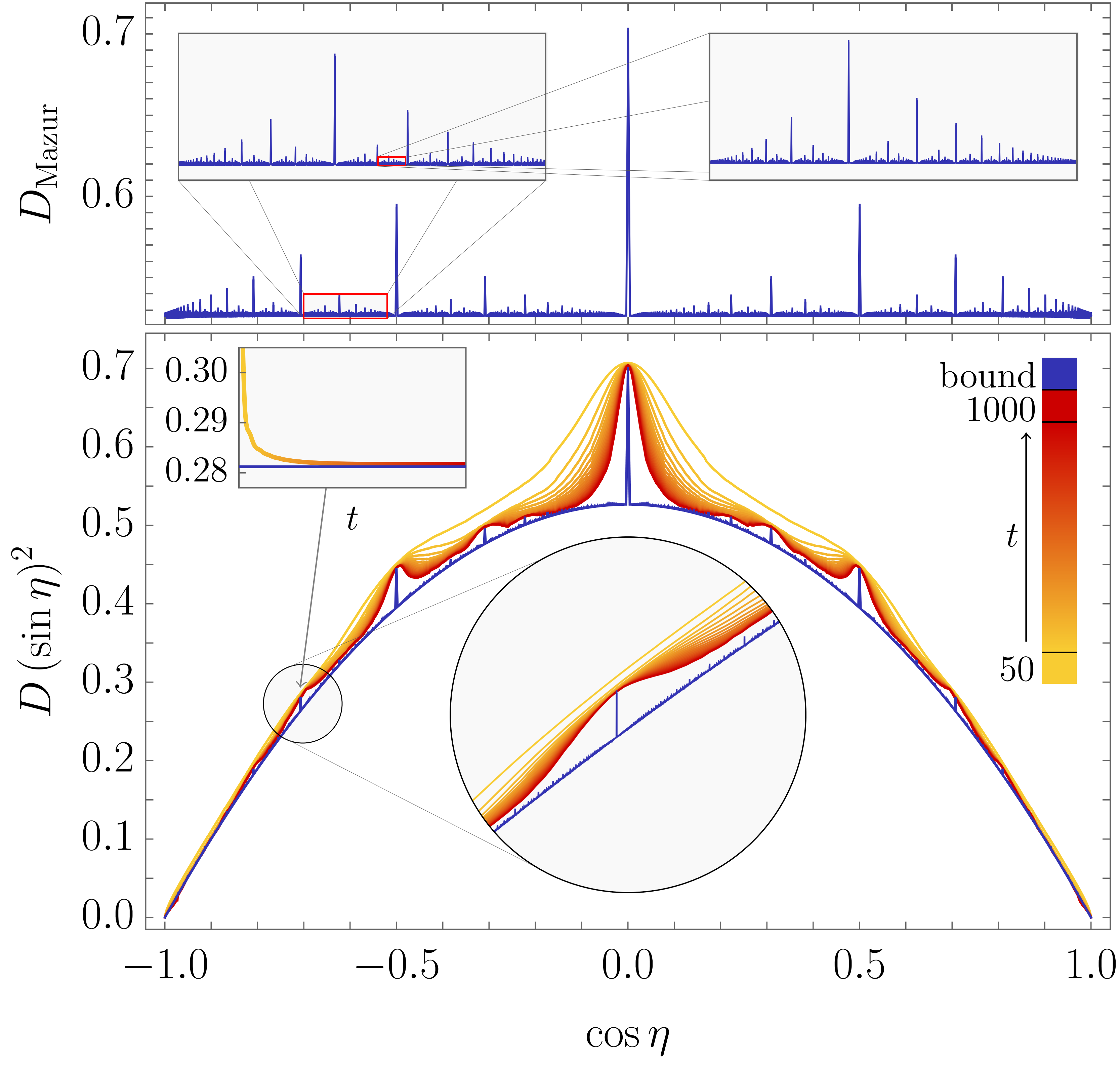}
        \caption{Drude weight at $|\lambda|=1$ as computed using the Mazur inequality (blue) and TEBD (yellow-red). The colour scales from yellow to red as the simulation time increases from $t=50$ to $t=1000$. The TEBD simulations were performed using a bond dimension of $64$ and a system size $N=3600$. The inset in the centre of the lower panel shows a more precise set of simulations using a bond dimension of $128$ for a small section of $\cos\eta$. The top-left inset of the lower panel shows convergence towards the fractal peak at $\eta=3\pi/4$ for bond dimension $256$. To demonstrate fractality, the upper panel only shows the Mazur bound without the rescaling.}
        \label{fig:drude}
    \end{figure}

\begin{figure}[ht!]
        \centering
        \hspace{-0.5cm}\includegraphics[width=0.9\linewidth]{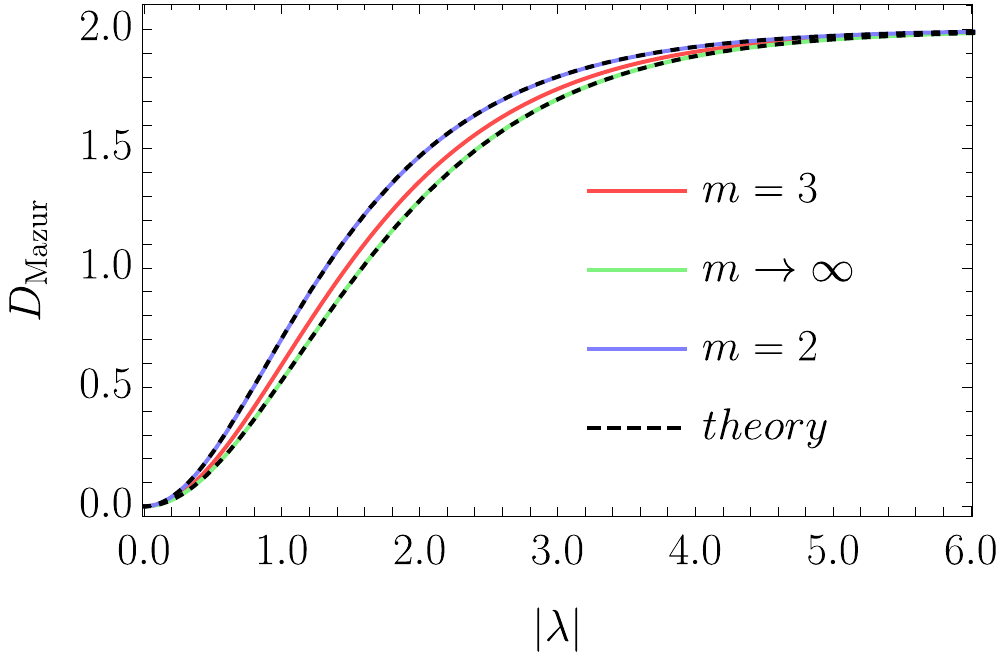}
        \caption{A comparison between the analytic and numerical results. Additionally we show $m=3$ case as an example of more generic behaviour with respect to $|\lambda|$. For all $m$, the $|\lambda|$-dependence lies between the $m=2$ and $m\to\infty$ curves.}
        \label{fig:compare}
    \end{figure}

    {\bf Mazur bound.--} 
    We now attempt to bound the spin Drude weight by means of the Mazur inequality as elaborated on in Ref.~\cite{tomaz14}. It can be rewritten in an integral form
    \begin{equation}
        D\ge D_{\rm Mazur}=\frac{1}{2}\,\real\int\d{^2\varphi}\,\overline{j(\bar{\varphi})}f(\varphi),\label{eq:mazur}
    \end{equation}
    where $f(\varphi)$ solves the following Fredholm equation
    \begin{equation}
        \int\d{^2\varphi'}\,K(\varphi,\varphi')f(\varphi')=j(\varphi).
        \label{eq:fredholm}
    \end{equation}
    The integrals are formally taken over the area of quasi-locality $|\real\,\varphi-\frac{\pi}{2}|<\frac{\pi}{2m}$. However, due to holomorphicity, a single line of integration centred at $\real\,\varphi=\frac{\pi}{2}$ is sufficient, which makes for an efficient quasi-exact numerical procedure of computing $D_{\rm Mazur}$. The full Mazur lower bound has a fractal dependence on $\eta$ and a continuous dependence on $|\lambda|$. It has been calculated numerically and compared to TEBD~\cite{vidal03,schollwock11} simulations -- see Appendix H of the SM~\cite{supp}. The dependence on $\eta$ is shown in Figure~\ref{fig:drude}, and the dependence on $|\lambda|$ in Figure~\ref{fig:compare}. 
    
    In Figure~\ref{fig:drude} we have re-scaled the Drude weight and the lower bound by a factor of $(\sin\eta)^2$ [see Eq.~\eqref{eq:hamiltonian}]. This allows us to make a comparison with the established continuous-time result~\cite{tomaz14}
    \begin{align}
        D'_{\rm Mazur}=\left[\frac{\sin\eta}{\sin(\pi/m)}\right]^2\left(1-\frac{m}{2\pi}\sin(2\pi/m)\right),
        \label{eq:continuous_bound}
    \end{align}
    found by expanding $(\sin\eta)^2\,D_{\rm Mazur}=\lambda^2\,D'_{\rm Mazur}+\mathcal{O}(\lambda^3)$ around $\lambda=0$. We can see this by noting, that the small-$\lambda$ expansion of Eqs.~\eqref{eq:mazur} and~\eqref{eq:fredholm} reproduces the corresponding equations in the continuous-time case.
    
    The integral equations can be solved analytically for $m\to\infty$, corresponding to an {\em irrational} value of $\eta/\pi$. The details are discussed in Appendix G of the SM~\cite{supp} and the result is the enveloping function 
    \begin{align}
        \lim_{m\to\infty}D_{\rm Mazur}=2\,\left(1-\frac{\Gd(|\lambda|)}{\sinh(|\lambda|)}\right),
        \label{eq:envelope_bound}
    \end{align}
    where $\Gd$ denotes the Gudermannian $\Gd(x)=2\arctan(e^x)-\pi/2$. This represents a continuous strict lower bound on top of which an additional fractal structure, shown in Figure~\ref{fig:drude} for $|\lambda|=1$, emerges.
    
    On the free fermion line ($\J'=0$ or $m=2$) exact diagonalization shows the saturation of the lower bound which includes only a single conserved quantity $Z(\pi/2)$. It can easily be computed to give
    \begin{align}
        D_{\rm Mazur}=2\,(1-{\rm sech}(|\lambda|)).
        \label{eq:bound_free}
    \end{align}
    For a complete $|\lambda|$ dependence of the Mazur lower bound see Figure~\ref{fig:compare}. Note that the $|\lambda|\to\infty$ limit is always $2$. This can easily be explained, since there, the local propagator reduces to a SWAP gate. As such, transport becomes perfectly ballistic with no scattering at all.
    
    {\bf Discussion.--} We have demonstrated and proven ballistic transport in a periodically driven interacting quantum spin chain, namely in the Trotterized $XXZ$ spin-1/2 model.
    We have used the notion of ballistic spin transport referring to a linearly growing extensive spin current after a quench from an inhomogeneous initial state with either a linear gradient or a step bias in the magnetization profile.
    We argue that this is the most natural definition of ballistic transport in the case of discrete-time propagation.
    Using the quasi-local conservation laws that we constructed by means of quantum-group theoretic methods, we have calculated the lower bound on the spin Drude weight and explicitly showed its fractal dependence on the anisotropy parameter. Extensive numerical simulations suggest the saturation of the lower bound -- see Figure S-2 in Appendix H of the SM. Note, however, that for a fixed commensurate anisotropy $\eta=l\,\pi/m$ the convergence with time seems to become extremely slow with increasing $m$, certainly beyond ultimate verification with state-of-the art numerical methods.
    
    In the continuous-time limit we correctly reproduce the well established results of ballistic spin transport in the $XXZ$ model. However, since the thermodynamic Bethe ansatz has not 
    yet been developed for driven integrable systems~\cite{vladimir}, our results open interesting new avenues for research.
    The conservation laws that we proposed (see also Ref.~\cite{zadnik}) can be directly applied for construction of complete generalized Gibbs ensembles and development of generalized hydrodynamics in integrable Floquet systems.
    
    \vspace{2mm}
    The authors thank M. \v Znidari\v c for helpful discussions on the topic. 
    The authors acknowledge support by the 
    European Research Council (ERC) through the advanced grant 694544 -- OMNES and the grant P1-0402 of Slovenian Research Agency (ARRS).

    \clearpage
    
    \onecolumngrid
    \appendix
    
    \setcounter{page}{1}
    \setcounter{equation}{0}
    \setcounter{figure}{0}
    
    \renewcommand{\theequation}{\textsc{S}-\arabic{equation}}
    \renewcommand{\thefigure}{\textsc{S}-\arabic{figure}}

    \begin{center}
        {\bfseries {\em Supplementary material for}\\ ``Ballistic spin transport in a periodically driven integrable quantum system''}     
    \end{center}
    
    \begin{center}
        Marko Ljubotina$^1$, Lenart Zadnik$^1$, Toma\v{z} Prosen$^1$\\
        $^1${\em Faculty of Mathematics and Physics, University of Ljubljana, Slovenia}
    \end{center}

    \section{Appendix A: Drude weight and Kubo formula}
    
    In this appendix we derive the Kubo form of the Drude weight starting from an initial state with a linear gradient of magnetization
     \begin{align}
        \rho_\mu=\frac{1}{2^N}\left(\mathbb{1}-\mu\sum_{n=1}^N\left(n-\frac{N+1}{2}\right)\sigma^\z_n\right).
    \end{align}
    Here $\mu$ is a perturbative parameter which corresponds to a local gradient in magnetization. Our goal is to examine the time dependent average $\langle J(t)\rangle_\mu=\tr{[\rho_\mu\,J(t)]}$ of the extensive current $J=\sum_n j_n$ for small $\mu$, where $J(t)=\U^{-t}\,J\,\U^{t}$. Note, that in our case the local current densities on even and odd sites are different. In the main text (MT) they are denoted by $j^{}_{2n}$ and $j'_{2n-1}$, respectively. Using the continuity equation $\U^\dagger\,\sigma_{n}^\z\,\U-\sigma_{n}^\z=-j_{n+1}+j_{n}$ (Eq.~\eqref{eq:scont} of the MT), which can be recast as 
    \begin{align}
        \U\,\sigma_{n}^\z\,\U^\dagger-\sigma_{n}^\z=\U\,(j_{n+1}-j_{n})\,\U^\dagger,
        \label{app:continuity}
    \end{align}
    we see, for an arbitrary observable $A$,
    \begin{align}
    \begin{gathered}
        \langle A(t)\rangle_\mu=\langle A\rangle-\frac{\mu}{2^N}\sum_{n=1}^N \left(n-\frac{N+1}{2}\right)\tr{[\U^t\sigma^z_n\,\U^{-t} A]}=\langle A\rangle-\frac{\mu}{2^N}\sum_{n=1}^N \left(n-\frac{N+1}{2}\right)\tr{[\U^{t-1}\sigma^z_n\,\U^{-t+1} A]}-\\
        -\frac{\mu}{2^N}\sum_{n=1}^N \left(n-\frac{N+1}{2}\right)\tr{[\U^{t}(j_{n+1}-j_{n})\,\U^{-t} A]}=\langle A(t-1)\rangle_\mu+\frac{\mu}{2^N}\sum_{n=1}^N\tr{[j_n\,\U^{-t}A\,\U^t]}=\langle A(t-1)\rangle_\mu+\mu\,\langle J A(t)\rangle.
    \end{gathered}
    \end{align}
    Using this recurrence we get
    \begin{align}
        \langle J(t)\rangle_\mu=\langle J\rangle_\mu+\mu\,\sum_{\tau=1}^t\,\langle J J(\tau)\rangle
        \label{app:Kubo1}
    \end{align}
    for the total extensive spin current $J$. Now, our definition of the Drude weight (Eq.~\eqref{eq:drude_weight} of the MT) gives the Kubo formula (Eq.~\eqref{eq:kubo} of the MT)
    \begin{align}
        D\equiv\lim_{t\to \infty}\lim_{N\to \infty}\lim_{\mu\to 0}\frac{\langle J(t)\rangle_\mu}{2 N t \mu}=\lim_{t\to \infty}\lim_{N\to \infty}\frac{1}{2N}\frac{1}{t}\sum_{\tau=1}^t\,\langle J J(\tau)\rangle.
        \label{app:Kubo2}
    \end{align}
    Taking the local current $j_n$ instead of $J$ in \eqref{app:Kubo1} we see 
    \begin{align}
    \lim_{t\to \infty}\lim_{N\to \infty}\lim_{\mu\to 0}\frac{\langle j_n(t)\rangle_\mu}{2 t \mu}=\lim_{t\to \infty}\lim_{N\to \infty}\frac{1}{2 t}\sum_{\tau=1}^t\,\langle J j_n(\tau)\rangle.
    \label{app:Kubo3}
    \end{align}
    Combining \eqref{app:Kubo2} and \eqref{app:Kubo3} for even and odd $n$ and using the invariance of the equilibrium average $\langle\bullet\rangle$ under the translation for two sites we now realize
    \begin{align}
       D=\lim_{t\to \infty}\lim_{N\to \infty}\lim_{\mu\to 0}\frac{1}{2 t \mu} \langle\frac{j_{2n}(t)+j'_{2n-1}(t)}{2}\rangle_\mu,
    \end{align}
    which shows that it is enough to measure the weight locally at any pair of neighbouring sites.
    
    \section{Appendix B: Integrability of the driven model}
    
    In this appendix we formulate the integrable structure of our driven model. Let $P$ denote the permutation operator which acts on the tensor square of the local physical space, $\V_p\otimes\V_p$, and has the explicit form $P_{n,n+1}=\frac{1}{2}\big(\mathbb{1}+\boldsymbol{\sigma}_n\cdot\boldsymbol{\sigma}_{n+1}\big)$, where $\boldsymbol{\sigma}=(\sigma^\x,\sigma^\y,\sigma^\z)$. Our local propagator, given by Eq.~\eqref{eq:propagator} of the MT, is a unitary matrix which can be written as $U=\check{R}(\lambda)=P R(\lambda)$ (see also Eq.~\eqref{eq:rmatrix} of the MT), where
    \begin{align}
        R(\lambda)=
        \left(\,
        \begin{matrix}
        1 & 0 & 0 & 0 \\
        0 & \frac{\sin\lambda}{\sin(\lambda+\eta)} & \frac{\sin\eta}{\sin(\lambda+\eta)} & 0 \\
        0 & \frac{\sin\eta}{\sin(\lambda+\eta)} & \frac{\sin\lambda}{\sin(\lambda+\eta)} & 0 \\
        0 & 0 & 0 & 1 \\
        \end{matrix}\,
        \right)\label{app:rmatrix}
    \end{align}
    satisfies the Yang-Baxter equation $R_{1,2}(\lambda-\mu)R_{1,3}(\lambda)R_{2,3}(\mu)=R_{2,3}(\mu)R_{1,3}(\lambda)R_{1,2}(\lambda-\mu)$ defined on the triple $\V_p\otimes\V_p\otimes\V_p$. This implies the commutation of transfer matrices $[T(\varphi),T(\varphi')]=0$,
    \begin{align}
        T(\varphi)=\tr{_0}\Big[\prod_{n=1}^{N}R_{n,0}(\varphi-(-1)^n\,\lhalf)\Big].\label{app:transfer1}
    \end{align}
    Index $0$ in \eqref{app:transfer1} denotes a copy of the physical vector space $\V_p$, which is traced out by the partial trace $\tr{_0}$. Elementary calculation yields 
    \begin{align}
        T(-\lhalf)^{-1}\,T(\lhalf)=\mathcal{U}_{odd}\mathcal{U}_{even}=\mathcal{U},
    \end{align}
        with $\mathcal{U}_{odd}$ and $\mathcal{U}_{even}$ defined in Eq.~\eqref{eq:propagation} of the MT. This establishes integrability, since $T(\varphi)$ generates local integrals of motion through logarithmic derivatives. For $\lambda=0$, the transfer matrix \eqref{app:transfer1} reduces to the one of the Heisenberg $XXZ$ model with the local Hamiltonian density
    \begin{align}
        h_{n,n+1}=\frac{1}{2\sin\eta}\big(\sigma_n^\x\sigma_{n+1}^\x+\sigma_n^\y\sigma_{n+1}^\y+\cos\eta\,(\sigma_n^\z\sigma_{n+1}^\z-\mathbb{1})\big).
    \end{align}

    \section{Appendix C: Transfer matrix from the complex spin representation}
    
    In this appendix we introduce the Lax operator and the transfer matrix which produces the spin-reversal asymmetric quasi-local integrals of motion $Z(\varphi)$. Such integrals exist only for $\eta=l\,\pi/m$, where $l$ and $m$ are co-prime integers. In our model $\lambda$ should then be purely imaginary. The starting point is the Lax operator of the $XXZ$ model,
     \begin{align}
        \L(\varphi,s)=\frac{1}{\sin\varphi}
        \begin{pmatrix}
        \sin(\varphi+\eta\,\S_s^\z) & \sin(\eta)\,\S_s^- \\
        \sin(\eta)\,\S_s^+ & \sin(\varphi-\eta\,\S_s^\z)
        \end{pmatrix}\label{app:lax}
    \end{align}
    which acts on the pair $\V_p\otimes\V_a$. $\V_a$ is an $m$-dimensional complex spin-$s$ representation of the quantum group $\mathcal{U}_q(sl_2)$, where $q=e^{i\eta}$. Its generators satisfy $[\S_s^+,\S_s^-]=\sin[2\eta\,\S_s^\z]/\sin\eta$, $[\S_s^\z,\S_s^\pm]=\pm\S_s^\pm$ and can take the following form
    \begin{align}
    \begin{aligned}
        &\S_s^\z=\sum_{k=0}^{m-1}(s-k)\ket{k}\bra{k},\\
        &\S_s^+=\sum_{k=0}^{m-2}\frac{\sin(k+1)\eta}{\sin\eta}\ket{k}\bra{k+1},\\
        &\S_s^-=\sum_{k=0}^{m-2}\frac{\sin(2s-k)\eta}{\sin\eta}\ket{k+1}\bra{k}.\label{app:Uqsl2}
    \end{aligned}
    \end{align}
  
    Along with the $R$-matrix \eqref{app:rmatrix}, the Lax operator \eqref{app:lax} satisfies another Yang-Baxter equation 
    \begin{align}
    R_{1,2}(\varphi-\varphi')\L_{1,a}(\varphi,s)\L_{2,a}(\varphi',s)=\L_{2,a}(\varphi',s)\L_{1,a}(\varphi,s)R_{1,2}(\varphi-\varphi')
    \label{app:ybe}
    \end{align}
    on the triple $\V_p\otimes\V_p\otimes\V_a$. As in Appendix B, we can use the Lax operator to construct a family of transfer matrices
    \begin{align}
      T(\varphi,s)=\tr{_a}\Big[\prod_{n=1}^N \L_{n,a}\big(\varphi-(-1)^n\,\lhalf,s\big)\Big].
      \label{app:transfer}
    \end{align}
    Yang-Baxter equations imply $[T(\varphi,s),T(\varphi',s)]=0$ and $[T(\varphi,s),T(\varphi')]=0$. As we will see, the spin-reversal asymmetric integrals of motion are
    \begin{align}
         Z(\varphi)=\frac{1}{2\eta\sin\eta}\,\partial_s T(\varphi,s)\mid_{s=0}.
         \label{app:integrals}
    \end{align}

    \section{Appendix D: Explicit form of the integrals of motion}
    
    Here we describe the explicit form of the conservation laws $Z(\varphi)$. Let us introduce the components of the Lax operator and its derivative at $s=0$ through
    \begin{align}
    \begin{aligned}
        \L(\varphi,0)=\sum_{\alpha\in\{0,+,-,\z\}}\sigma^\alpha\otimes\L^\alpha(\varphi),\qquad
        \partial_s \L(\varphi,s)\mid_{s=0}=\sum_{\alpha\in\{0,+,-,\z\}}\sigma^\alpha\otimes\widetilde{\L}^\alpha(\varphi),
        \label{app:components1}
    \end{aligned}
    \end{align} 
    where $\sigma^0=\mathbb{1}$ denotes the identity. Components $\L^\alpha$, $\widetilde{\L}^\alpha$ act on the auxiliary space $\V_a$ and explicitly read
    \begin{align}
    \begin{aligned}
        &\L^0(\varphi)=\sum_{k=0}^{m-1}\cos (k\eta) \ket{k}\bra{k},&\qquad &\widetilde{\L}^0(\varphi)=\eta\sum_{k=1}^{m-1}\sin (k\eta) \ket{k}\bra{k},\\
        &\L^+(\varphi)=-\frac{1}{\sin\varphi}\sum_{k=1}^{m-2}\sin (k\eta) \ket{k+1}\bra{k},& &\widetilde{\L}^+(\varphi)=\frac{2\eta}{\sin\varphi}\sum_{k=0}^{m-2}\cos (k\eta) \ket{k+1}\bra{k},\\
        &\L^-(\varphi)=\frac{1}{\sin\varphi}\sum_{k=0}^{m-2}\sin [(k+1)\eta] \ket{k}\bra{k+1},&  &\widetilde{\L}^-(\varphi)=0,\\
        &\L^\z(\varphi)=-\cot\varphi\sum_{k=1}^{m-1}\sin (k\eta) \ket{k}\bra{k},& &\widetilde{\L}^\z(\varphi)=\eta\cot\varphi\sum_{k=0}^{m-1}\cos (k\eta) \ket{k}\bra{k}.
        \label{app:components2}
    \end{aligned}
    \end{align}
    We should now make the following observations:
    \begin{enumerate}
        \item Of the derivative components, only $\widetilde{\L}^\z(\varphi)$ preserves the highest-weight state $\ket{0}$.
        \item Of the Lax components, only $\L^0(\varphi)$ preserves the highest-weight state $\ket{0}$.
        \item No Lax component $\L^\alpha(\varphi)$, where $\alpha\in\{0,+,-,\z\}$ can lift us from the {\em highest-weight} state $\ket{0}$ into the {\em reduced subspace} $\V_a'={\rm lsp}\{\ket{1},\ket{2},...,\ket{m-1}\}$.
    \end{enumerate}
    Because of the partial trace over the auxiliary space in \eqref{app:transfer}, the derivative with respect to $s$ in \eqref{app:integrals} can always be shifted to the rightmost position in the string of Lax operators. The trace can then be separated as $\tr{_a}(\bullet)=\bra{0}\bullet\ket{0}+\sum_{k=1}^{m-1}\bra{k}\bullet\ket{k}$. As we will see, the first term will produce a linearly extensive contribution and the second term a remainder which is exponentially small in the system's size.
    
    Suppressing the $\varphi$-dependence, let us first focus on the string $\bra{0}\L^{\alpha_1}\L^{\alpha_2}...\widetilde{\L}^{\alpha_N}\ket{0}$. Due to the first two observations, if $\alpha_N=\z$ all $\alpha_n$ for $n<N$ must be zero. This produces the following magnetization-like terms
    \begin{align}
        \sum_{n=0}^{N/2-1}\frac{\cot(\varphi-\lambda/2)}{2\sin\eta}\,\sigma^\z_{2n}+\frac{\cot(\varphi+\lambda/2)}{2\sin\eta}\,\sigma^\z_{2n+1}.
        \label{app:mag_terms}
    \end{align}
    If $\alpha_N=+$, the only option is to have $\alpha_L=-$ for some fixed $L<N$ and $\alpha_n=0$ for all $n<L$. This produces the following terms 
    \begin{align}
    \begin{gathered}
      \sum_{n=0}^{N/2-1}S^{2n}\Big(\sum_{r=1}^{N/2}q_{2r}^-(\varphi)+\sum_{r=1}^{N/2-1}q_{2r+1}^-(\varphi)\Big)
      +S^{2n+1}\Big(\sum_{r=1}^{N/2}q_{2r}^+(\varphi)+\sum_{r=1}^{N/2-1}q_{2r+1}^+(\varphi)\Big),\label{app:ql_terms}
    \end{gathered}
    \end{align}
    where $S^{n}$ is an automorphism of operator algebra implementing a periodic shift for $n$ lattice sites, defined by $S^n(\sigma_{L}^\alpha)=\sigma^\alpha_{L+n}$, and 
    \begin{align}
    \begin{aligned}
      &q_{2r}^\pm(\varphi)=\sum_{\alpha_1,...,\alpha_{2r-2}}\frac{\bra{1}\L^{\alpha_1}(\varphi\pm\lhalf)...\L^{\alpha_{2r-2}}(\varphi\mp\lhalf)\ket{1}}{\sin(\varphi+\lhalf)\sin(\varphi-\lhalf)}\,\sigma^-\otimes\sigma^{\alpha_1}\otimes...\otimes\sigma^{\alpha_{2r-2}}\otimes\sigma^+\otimes\mathbb{1}^{\otimes (N-2r)},\\[1em]
      &q_{2r+1}^\pm(\varphi)=\sum_{\alpha_1,...,\alpha_{2r-1}}\frac{\bra{1}\L^{\alpha_1}(\varphi\pm\lhalf)...\L^{\alpha_{2r-1}}(\varphi\pm\lhalf)\ket{1}}{[\sin(\varphi\mp\lhalf)]^2}\,\sigma^-\otimes\sigma^{\alpha_1}\otimes...\otimes\sigma^{\alpha_{2r-1}}\otimes\sigma^+\otimes\mathbb{1}^{\otimes (N-2r-1)}.\label{app:densities}
    \end{aligned}
    \end{align}
    
    The rest of the trace, $\sum_{k=1}^{m-1}\bra{k}\L^{\alpha_1}\L^{\alpha_2}...\widetilde{\L}^{\alpha_N}\ket{k}$, will produce the remainder. Due to the third observation it can be rewritten as
    \begin{align}
        p_N(\varphi)=\partial_s\left( \tr{'_a}\Big[\prod_{n=1}^{N}\Big(\sum_{k'=1}^{m-1}\ket{k'}\bra{k'}_a\Big)\L_{n,a}\big(\varphi-(-1)^n\,\lhalf,s\big)\Big(\sum_{k=1}^{m-1}\ket{k}\bra{k}_a\Big)\Big]\right)_{s=0},
        \label{app:remainder}
    \end{align}
    where $\tr{'_a}$ denotes the partial trace over the reduced subspace $\V_a'={\rm lsp}\{\ket{1},\ket{2},...,\ket{m-1}\}$ onto which the Lax components have been projected by $\sum_{k=1}^{m-1}\ket{k}\bra{k}_a$. Gathering all of the terms with explicit matrix product forms provided by equations \eqref{app:densities} and \eqref{app:remainder} we have 
    \begin{align}
    \begin{gathered}
      Z(\varphi)=Z_\infty(\varphi)+p_N(\varphi),\\
      Z_\infty(\varphi)=\sum_{n=0}^{N/2-1}\Big\{S^{2n}\Big(\sum_{r=1}^{N/2}q_{2r}^-(\varphi)+\sum_{r=1}^{N/2-1}q_{2r+1}^-(\varphi)\Big)
      +S^{2n+1}\Big(\sum_{r=1}^{N/2}q_{2r}^+(\varphi)+\sum_{r=1}^{N/2-1}q_{2r+1}^+(\varphi)\Big)+\\[1em]
      +\frac{\cot(\varphi-\lambda/2)}{2\sin\eta}\,\sigma^\z_{2n}+\frac{\cot(\varphi+\lambda/2)}{2\sin\eta}\,\sigma^\z_{2n+1}\Big\}.
      \label{app:explicit}
    \end{gathered}
    \end{align}
    By $Z_\infty(\varphi)$ we have denoted the part of $Z(\varphi)$ that survives the thermodynamic limit $N\to\infty$, as we will see in Appendix F.

    \section{Appendix E: Spin-reversal symmetric and anti-symmetric component}
    
    In this appendix we define the decomposition of the integrals of motion $Z(\varphi)$ with respect to the symmetry under the spin-reversal $\mathcal{P}=\prod_{n=1}^N\sigma_n^\x$. Let us assume $\lim_{N\to 0}p_N(\varphi)=0$ in \eqref{app:explicit}. The spin-reversal symmetric and anti-symmetric components of $Z(\varphi)$ are defined as 
    \begin{align}
    Z^\pm(\varphi)=\frac{1}{2}(Z(\varphi)\pm \mathcal{P}Z(\varphi)\mathcal{P}).
    \label{app:parity_decomp}
    \end{align}
    Note, that the Lax operator \eqref{app:lax} satisfies 
    $\sigma_n^\x\L_{n,a}^\alpha(\varphi,s)\sigma_n^\x=\L_{n,a}^\alpha(\pi-\varphi,s)^T$,
    where $(\bullet)^T$ denotes the partial transposition with respect to the physical space. This immediately implies
    \begin{align}
        \mathcal{P}Z(\varphi)\mathcal{P}=Z(\pi-\varphi)^T\mid_{\lambda\to-\lambda}.\label{app:parity_transform}
    \end{align}
    Since the operator $Z(\varphi)$ is not symmetric [see Eq.~\eqref{app:densities}] the spin-reversal anti-symmetric component $Z^-(\varphi)$ is nonzero. This is crucial, since it provides the overlap with the spin current.
    
    We can now make these components orthogonal with respect to the total magnetization $M=\sum_{n=1}^N\sigma_n^\z$ which does not contribute to the overlap with the spin current. We have
    \begin{align}
      Z_\perp^{+}(\varphi)=Z^{+}(\varphi),\qquad
      Z_\perp^{-}(\varphi)=Z^{-}(\varphi)-\frac{1}{N}\langle M,Z(\varphi)\rangle\,M,\label{app:orthogonalized_charges}
    \end{align}
    with
    \begin{align}
        \lim_{N\to\infty}\frac{1}{N}\langle M,Z(\varphi)\rangle=\frac{\sin2\varphi}{2\sin\eta\,(\cos\lambda-\cos2\varphi)}.\label{app:magnetization_overlap}
    \end{align}
   
    \section{Appendix F: Quasi-locality and the inner product between the integrals of motion}
    
\begin{figure}[ht!]
        \centering
        \includegraphics[width=500pt]{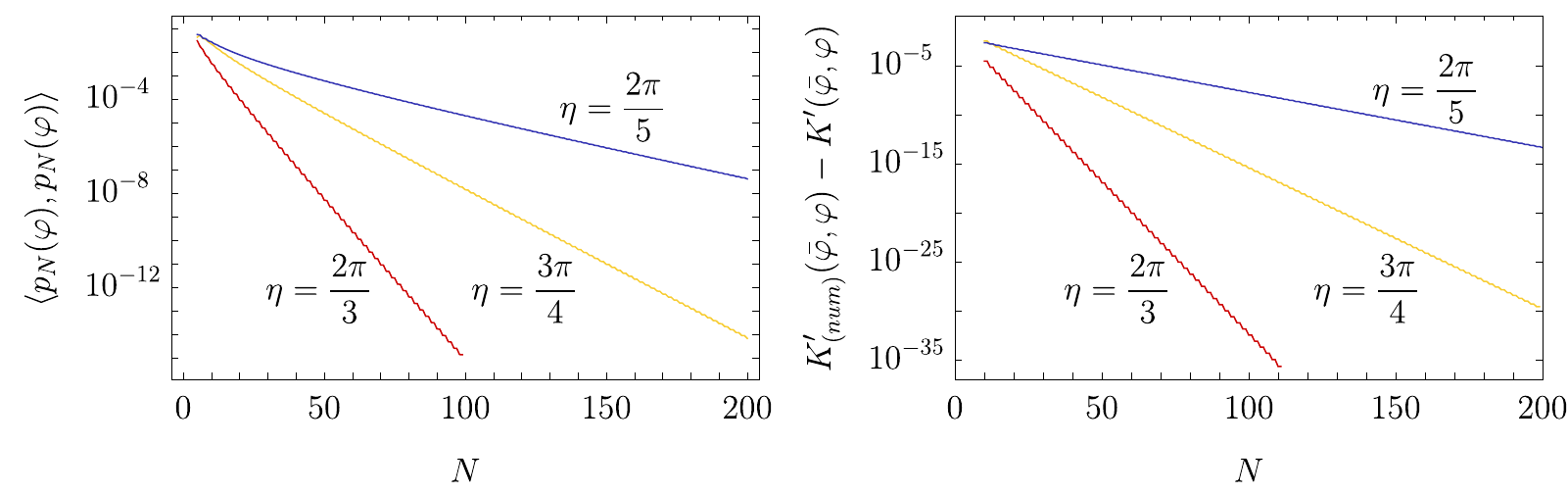}
        \caption{On the left, the norm $\langle p_N(\varphi),p_N(\varphi)\rangle$ of the remainder at $\lambda=i\frac{3}{4}$ and $\varphi=\frac{\pi}{2}+i\frac{3}{5}$ is shown. The right diagram shows the difference between numerically computed $K'(\bar{\varphi},\varphi)$ \eqref{app:norm3} and our conjecture \eqref{app:conjecture}, which is finite in the thermodynamic limit.}
        \label{fig:remainder}
    \end{figure}

    This appendix describes the quasi-locality of $Z(\varphi)$ given in \eqref{app:explicit} and provides explicit formulas for their overlaps. We will refer to operators $Z(\varphi)$ as {\em quasi-local} if $\langle Z(\varphi),Z(\varphi)\rangle\propto N$ for large $N$, where $\langle A,B\rangle=\tr{(A^\dagger B)}/2^N$ denotes the Hilbert-Schmidt inner product~\cite{tomaz14}. We will see that the following holds:
    \begin{prop}
    Let $\lambda$ be purely imaginary and $\eta=\frac{l\,\pi}{m}$ with $l$ and $m$ co-prime integers. Operators $Z(\varphi)$ in \eqref{app:explicit} with terms given by \eqref{app:densities} and \eqref{app:remainder} are quasi-local if $|\real\,\varphi-\frac{\pi}{2}|<\frac{\pi}{2m}$.
    \label{prop:prop1}
    \end{prop}
    \noindent We start by recalling the decomposition $Z(\varphi)=Z_\infty(\varphi)+p_N(\varphi)$ from \eqref{app:explicit} and writing the inner product
    \begin{align}
        \lim_{N\to\infty}\frac{1}{N}\langle Z(\bar{\varphi}),Z(\varphi')\rangle=\lim_{N\to\infty}\frac{1}{N}\Big(\langle Z_\infty(\bar{\varphi}),Z_\infty(\varphi')\rangle+\langle Z_\infty(\bar{\varphi}),p_N(\varphi')\rangle+\langle p_N(\bar{\varphi}),Z_\infty(\varphi')\rangle+\langle p_N(\bar{\varphi}),p_N(\varphi')\rangle\Big),
        \label{app:norm1}
    \end{align}
    where $(\bar{\bullet})$ denotes the complex conjugation. The first term can be written as
    \begin{align}
    \begin{gathered}
        \lim_{N\to\infty}\frac{1}{N}\langle Z_\infty(\bar{\varphi}),Z_\infty(\varphi')\rangle=K'(\varphi,\varphi')
        +\frac{(\sin\lambda)^2-\sin2\varphi\,\sin2\varphi'}{4\, (\sin\eta)^2\,(\cos2\varphi-\cos\lambda)\,(\cos\lambda-\cos2\varphi')},
        \label{app:norm2}
    \end{gathered}
    \end{align}
    where
    \begin{align}
    \begin{gathered}
        K'(\varphi,\varphi')=\frac{1}{2}\sum_{r=1}^\infty\Big(\langle q_{2r}^-(\bar{\varphi}),q_{2r}^-(\varphi')\rangle+\langle q_{2r+1}^-(\bar{\varphi}),q_{2r+1}^-(\varphi')\rangle+\langle q_{2r}^+(\bar{\varphi}),q_{2r}^+(\varphi')\rangle+\langle q_{2r+1}^+(\bar{\varphi}),q_{2r+1}^+(\varphi')\rangle\Big).
        \label{app:norm3}
    \end{gathered}
    \end{align}
    Using the Cauchy-Schwarz inequality, the absolute value of the last three terms in \eqref{app:norm1} can be bounded by the Hilbert-Schmidt norms of $Z_\infty(\varphi)$ and $p_N(\varphi)$. Proposition \ref{prop:prop1} then holds if the following is true:
    \begin{lema}
    The Hilbert-Schmidt norm of the remainder $p_N(\varphi)$ is exponentially small in the system size $N$ and $K'(\bar{\varphi},\varphi)$ is finite if $|\real\,\varphi-\frac{\pi}{2}|<\frac{\pi}{2m}$.
    \end{lema}
    \noindent Figure~\ref{fig:remainder} shows the exponential suppression of the norm $\langle p_N(\varphi),p_N(\varphi)\rangle$ of the remainder (left) and the difference $K_{(num)}'(\bar{\varphi},\varphi)-K'(\bar{\varphi},\varphi)$ between the numerically computed $K'(\bar{\varphi},\varphi)$ \eqref{app:norm3} and our conjectured formula \eqref{app:conjecture} which is finite in the thermodynamic limit (right).
   
   In the following we describe how to calculate $K'(\varphi,\varphi')$ and conjecture its explicit formula, which we also use in the MT. To this end, let us define a transfer matrix $\mathbb{T}(\varphi,\varphi')=\sum_{\alpha\in\{0,+,-,\z\}}\L^\alpha(\varphi)\otimes\L^\alpha(\varphi')\,\langle\sigma^\alpha,\sigma^\alpha\rangle$ over ${\cal V}_a\otimes {\cal V}_a$. This enables us to write down the inner products of the local densities in a computationally convenient form
    \begin{align}
    \begin{aligned}
        &\langle q_{2r}^\pm(\bar{\varphi}),q_{2r}^\pm(\varphi')\rangle=\frac{\bra{1}\otimes\bra{1}\big[\mathbb{T}(\varphi\mp\lhalf,\varphi'\pm\lhalf)\mathbb{T}(\varphi\pm\lhalf,\varphi'\mp\lhalf)\big]^{r-1}\ket{1}\otimes\ket{1}}{4\,\sin(\varphi-\lhalf)\sin(\varphi+\lhalf)\sin(\varphi'+\lhalf)\sin(\varphi'-\lhalf)},\\
        &\langle q_{2r+1}^\pm(\bar{\varphi}),q_{2r+1}^\pm(\varphi')\rangle=\frac{\bra{1}\otimes\bra{1}\mathbb{T}(\varphi\mp\lhalf,\varphi'\pm\lhalf)\big[\mathbb{T}(\varphi\pm\lhalf,\varphi'\mp\lhalf)\mathbb{T}(\varphi\mp\lhalf,\varphi'\pm\lhalf)\big]^{r-1}\ket{1}\otimes\ket{1}}{4\,[\sin(\varphi\pm\lhalf)\sin(\varphi'\mp\lhalf)]^2}.
        \label{app:density_norm}
    \end{aligned}
    \end{align}
    Calculating the matrix $\mathbb{T}(\varphi,\varphi')$ we see, that it 
    \begin{enumerate}
        \item preserves the vector $\ket{0}\otimes\ket{0}$ and
        \item preserves the subspace spanned by the vectors of the form $\ket{k}\otimes\ket{k}$, $k=0,1,...,m-1$.
    \end{enumerate} 
    Due to the first property, any excursion from the reduced auxiliary space $\V_a'={\rm lsp}\{\ket{1},...,\ket{m-1}\}$ in the string of matrices $\mathbb{T}(\varphi,\varphi')$ gives zero.
    This allows us to project these matrices onto the reduced auxiliary subspace $\V_a'$ by identifying $\ket{k}\otimes\ket{k}\longleftrightarrow |\sin (k\eta)|\ket{k}$ and $\bra{k}\otimes\bra{k}\longleftrightarrow |\sin (k\eta)|^{-1}\bra{k}$. The matrix $\mathbb{T}(\varphi,\varphi')$ becomes
    \begin{align}
        \T(\varphi,\varphi')=\sum_{k=1}^{m-1}\left([\cos (k\eta)]^2+\cot\varphi\cot\varphi' [\sin (k\eta)]^2\right)\ket{k}\bra{k}+\frac{|\sin (k\eta)\,\sin [(k+1)\eta]|}{2\sin\varphi\sin\varphi'}\big(\ket{k}\bra{k+1}+\ket{k+1}\bra{k}\big).
        \label{app:aux_transfer}
    \end{align}
    This matrix has two important properties:
    \begin{enumerate}
        \item For $|\real\,\varphi-\frac{\pi}{2}|<\frac{\pi}{2m}$ its eigenvalues are strictly below $1$ in absolute value. This has been proven in~\cite{tomaz14}. Obviously this also holds for $\T(\varphi\pm\lhalf,\varphi'\mp\lhalf)$ if $\lambda$ is strictly imaginary.
        \item Commutation $[\T(\varphi+\lhalf,\varphi'-\lhalf),\T(\varphi+\mhalf,\varphi'-\mhalf)]=0$, which is actually a result of the Yang-Baxter equation~\eqref{app:ybe} and its transpose.
    \end{enumerate}
    Using these two properties we can sum up the series over $r$ in \eqref{app:norm3} to get
    \begin{align}
        K'(\varphi,\varphi')=\frac{1}{8}\bra{1}\big(2+\frac{\sin(\varphi+\lhalf)\sin(\varphi'-\lhalf)}{\sin(\varphi-\lhalf)\sin(\varphi'+\lhalf)}\,\T(\varphi+\lhalf,\varphi'-\lhalf)+\frac{\sin(\varphi-\lhalf)\sin(\varphi'+\lhalf)}{\sin(\varphi+\lhalf)\sin(\varphi'-\lhalf)}\,\T(\varphi-\lhalf,\varphi'+\lhalf)\big)\ket{\psi},
        \label{app:kernel}
    \end{align}
    where $\ket{\psi}=\sum_{k=1}^{m-1}\psi_k\ket{k}$ is a solution of a non-homogeneous recurrence relation of the fourth order with non-constant coefficients,
    \begin{align}
        \sin(\varphi+\lhalf)\sin(\varphi'+\lhalf)\sin(\varphi-\lhalf)\sin(\varphi'-\lhalf)\,\Big[1-\T(\varphi-\lhalf,\varphi'+\lhalf)\T(\varphi+\lhalf,\varphi'-\lhalf)\Big]\ket{\psi}=\ket{1}.
        \label{app:recurrence}
    \end{align}
    In order to compute the overlaps explicitly, a solution of the recurrence relation \eqref{app:recurrence} is now needed. Up to now we haven't been able to solve it explicitly, however, we have guessed and extensively numerically checked the following:
    \begin{conj} 
    Let $\lambda$ be purely imaginary and $\eta=\frac{l\,\pi}{m}$, $l,\,m$ co-prime integers. Sum \eqref{app:norm3} of the overlaps between the local densities of $Z(\varphi)$ is
    \begin{align}
        K'(\varphi,\varphi')=\frac{\big(\cos(\varphi-\varphi'+\lambda)+\cos(\varphi-\varphi'-\lambda)-2\cos(\varphi+\varphi')\big)\,\sin((m-1)(\varphi+\varphi'))}{2\,(\sin\eta)^2\,(\cos2\varphi-\cos\lambda)\,(\cos\lambda-\cos2\varphi')\,\sin(m(\varphi+\varphi'))}.
        \label{app:conjecture}
    \end{align}
    \end{conj}
    \noindent For small $\lambda$ this expression correctly reproduces the continuous-time result -- see~\cite{tomaz14}.
    Since the remainder $p_N(\varphi)$ in the conservation laws $Z(\varphi)$ vanishes in the thermodynamic limit $N\to\infty$ we have
    \begin{align}
        \lim_{N\to\infty}\frac{1}{N}\langle Z(\bar{\varphi}),Z(\varphi')\rangle=\lim_{N\to\infty}\frac{1}{N}\langle Z_\infty(\bar{\varphi}),Z_\infty(\varphi')\rangle
        \label{app:thermodynamic}
    \end{align} 
    and the full overlap between the conservation laws can be computed from \eqref{app:norm2}.
    Using \eqref{app:parity_transform} we also deduce
    \begin{align}
        \lim_{N\to\infty}\frac{1}{N}\langle Z(\bar{\varphi}),{\cal P}Z(\varphi){\cal P}\rangle=-\frac{(\sin\lambda)^2-\sin2\varphi\,\sin2\varphi'}{4\, (\sin\eta)^2\,(\cos2\varphi-\cos\lambda)\,(\cos\lambda-\cos2\varphi')}.
    \end{align}
    It is now easy to compute the overlaps between the spin-reversal anti-symmetric components $Z^-(\varphi)$ given in \eqref{app:parity_decomp} and, separately, the magnetization-orthogonal components $Z_\perp^-(\varphi)$ given in \eqref{app:orthogonalized_charges}:
    \begin{align}
    \begin{aligned}
        &\lim_{N\to\infty}\frac{1}{N}\langle Z^-(\bar{\varphi}),Z^-(\varphi')\rangle=\frac{1}{2}K'(\varphi,\varphi')+\frac{(\sin\lambda)^2-\sin2\varphi\,\sin2\varphi'}{4\, (\sin\eta)^2\,(\cos2\varphi-\cos\lambda)\,(\cos\lambda-\cos2\varphi')},\\
        &\lim_{N\to\infty}\frac{1}{N}\langle Z_\perp^-(\bar{\varphi}),Z_\perp^-(\varphi')\rangle=\frac{1}{2}K'(\varphi,\varphi')
        +\frac{(\sin\lambda)^2}{4(\sin\eta)^2\,(\cos2\varphi-\cos\lambda)\,(\cos\lambda-\cos2\varphi')}.
        \label{app:HS_kernel}
   \end{aligned}
   \end{align}
   The last overlap is essentially Eq.~\eqref{eq:norms} of the MT and is used in the Mazur inequality. Therefore, it deserves its own symbol  $K(\varphi,\varphi')\equiv\lim_{N\to\infty}\frac{1}{N}\langle Z_\perp^-(\bar{\varphi}),Z_\perp^-(\varphi')\rangle$. Explicitly it reads
   \begin{align}
       K(\varphi,\varphi')=\frac{\big(\cos(\varphi-\varphi'+\lambda)+\cos(\varphi-\varphi'-\lambda)-2\cos(\varphi+\varphi')\big)\,\sin((m-1)(\varphi+\varphi'))+(\sin\lambda)^2\,\sin(m(\varphi+\varphi'))}{4\,(\sin\eta)^2\,(\cos2\varphi-\cos\lambda)\,(\cos\lambda-\cos2\varphi')\,\sin(m(\varphi+\varphi'))}.
        \label{app:total_kernel}
   \end{align}

   \section{Appendix G: Solving the integral equation}
   
   In this appendix we describe how to tackle the integral equation given by Eq.~\eqref{eq:fredholm} of the MT. Numerics suggests that it is enough to restrict the domain of integration to the line $\rm{Re}\,\varphi=\pi/2$, which can be understood in terms of analyticity of the kernel. Taking $\varphi=\pi/2+i\log x$, $\varphi'=\pi/2+i\log x'$ and $\lambda=i|\lambda|=i\log\Lambda$ we see, that the integral equation and the lower bound (Eqs.~\eqref{eq:fredholm} and~\eqref{eq:mazur} of the MT, respectively) are equivalent to 
   \begin{align}
   \int_0^\infty\d{x'}\,\widetilde{K}(x,x')\widetilde{f}(x')=1,\qquad D\ge{\rm Re}\int_0^\infty\d{x}\,\widetilde{f}(x)\widetilde{j}(x),
   \label{app:fredholm0}
   \end{align}
   where
   \begin{align}
       x\,\widetilde{f}(x)=\left[\frac{f(\varphi)}{(\sin\eta\,\sin\lambda)}\right]_{\varphi=\pi/2+i\log x,\,\lambda=i\log\Lambda},\qquad
       \widetilde{j}(x)=\left[\frac{\sin\eta\,\sin\lambda}{2}\,\overline{j(\bar{\varphi})}\right]_{\varphi=\pi/2+i\log x,\,\lambda=i\log\Lambda}
    \label{app:fredholm1}
   \end{align}
   and
   \begin{align}
       \widetilde{K}(x,x')=\left[\frac{K(\varphi,\varphi')}{j(\varphi)}\,(\sin\eta\,\sin\lambda)\right]_{\varphi=\pi/2+i\log x,\,\varphi'=\pi/2+i\log x',\,\lambda=i\log\Lambda}.
       \label{app:fredholm2}
   \end{align}
   $K(\varphi,\varphi')$ is given by~\eqref{app:total_kernel}, while Eq.~\eqref{eq:overlap} of the MT provides explicit formula for $j(\varphi)$. Function $\widetilde{f}(x)$ has absorbed the Jacobian $1/x$ of the coordinate transformation $\log x\to x$. The goal is now to solve for $\widetilde{f}(x)$ and plug it into the lower bound, provided the explicit form of real functions $\widetilde{K}(x,x')$ and $\widetilde{j}(x)$
   \begin{align} 
   \begin{gathered}
   \widetilde{K}(x,x')=\frac{\left(2 \Lambda +\left(\Lambda ^2+1\right) x^2\right) \left(2 \Lambda
   +\left(\Lambda ^2+1\right) x'^2\right) (x x')^{2 m}-x^2 x'^2 \left(\Lambda ^2+2 \Lambda  x^2+1\right) \left(\Lambda ^2+2 \Lambda  x'^2+1\right)}{8 \Lambda  x^2 \left(\Lambda +x'^2\right) \left(\Lambda  x'^2+1\right) \left((x x')^{2 m}-1\right)},\\
   \widetilde{j}(x)=\frac{\left(\Lambda ^2-1\right)^2 x^2}{4 \Lambda  \left(\Lambda +x^2\right) \left(\Lambda  x^2+1\right)}.
   \label{app:fredholm3}
   \end{gathered}
   \end{align}
   Since $\widetilde{j}(x)=\widetilde{j}(x^{-1})$, changing the variables according to $x\to1/x$ in the second equation of \eqref{app:fredholm0} suggests $\widetilde{f}(x^{-1})=x^2\widetilde{f}(x)$. Using this symmetry, we can rewrite the first equation of \eqref{app:fredholm0} as
   \begin{align}
       \left(1+2\Lambda x^2+\Lambda^2\right) G(x^{-1})+\left(1+2\Lambda x^{-2}+\Lambda^2\right) G(x)=1, 
       \label{app:fredholm4}
   \end{align}
   where
   \begin{align}
       G(x)=\int_0^\infty\d{x'}\frac{\widetilde{f}(x')L(x')}{1-(x'/x)^{2m}},\qquad L(x)=\frac{x^2(1+2\Lambda x^2+\Lambda^2)}{8\Lambda(\Lambda+x^2)(1+\Lambda x^2)}.
       \label{app:fredholm5}
   \end{align}
   Taking the limit $m\to\infty$ now gives 
   \begin{align}
       G(x)=\int_0^x \d{x'} \widetilde{f}(x')L(x'),\qquad \widetilde{f}(x)=\frac{G'(x)}{L(x)}.
       \label{app:fredholm6}
   \end{align}
   If we define $g(x)=(1+2\Lambda x^{-2}+\Lambda^2)G(x)$, we have $g(x)+g(x^{-1})=1$. So we can assume the following expansion
   \begin{align}
       g(x)=\frac{1}{2}+\sum_{m=1}^\infty c_m \frac{x^m-x^{-m}}{2},
       \label{app:fredholm7}
   \end{align}
   for some coefficients $c_m$. If we use this in the second equation of \eqref{app:fredholm6} we see that $G(x)=x^4 G(x^{-1})$
   must hold for $\widetilde{f}(x^{-1})=x^2\widetilde{f}(x)$ to be true. Together with \eqref{app:fredholm4} this is enough to solve for $G$ and then $\widetilde{f}$, which produces
   \begin{align}
       \widetilde{f}(x)=\frac{32 \Lambda  x \left(\Lambda +x^2\right) \left(\Lambda  x^2+1\right)}{\left(\Lambda ^2+\left(\Lambda ^2+1\right) x^4+4 \Lambda x^2+1\right)^2}.
       \label{app:fredholm8}
   \end{align}
   It is now easy to compute the lower bound \eqref{app:fredholm0} which, after substitution $\Lambda=e^{|\lambda|}$ becomes
   \begin{align}
       D\ge 2\,\left(1-\frac{\Gd(|\lambda|)}{\sinh(|\lambda|)}\right).
   \end{align}
   This is the enveloping function given in Eq.~\eqref{eq:envelope_bound} of the MT. $\Gd$ denotes the Gudermannian function with explicit formula $\Gd(x)=2\arctan(e^x)-\pi/2$.
   
    \section{Appendix H: Numerical approaches}
    
    \textbf{Mazur bound calculation.--}
    In order to compute the Drude weight bound numerically (for general $m$), we discretize the region of integration ($|\real\,\varphi-\frac{\pi}{2}|<\frac{\pi}{2m}$) into $L$ equally-sized rectangles and choose a cut-off value $|\imag\,\varphi|\le\mathcal{C}$. In our experience cut-offs $\mathcal{C}=5$ and $\mathcal{C}=10$ are already indistinguishable and thus the first one ($\mathcal{C}=5$) is sufficient for accurate results. It also appears we can limit ourselves to the line $\real\varphi=\frac{\pi}{2}$ as increasing the number of points in the real direction does not affect the value significantly. 
    
    Having discretized the region, the integral equation (Eq.~\eqref{eq:fredholm} of the MT) reduces to a system of linear equations which can be written in matrix form
    \begin{equation}
        \boldsymbol{K}\cdot\boldsymbol{f}=\boldsymbol{j}
        \label{app:num1}
    \end{equation}
    and solved efficiently for $\boldsymbol{f}$. Here, $\boldsymbol{K}$ represents a square matrix with elements $K(\varphi,\varphi')$ which are computed from \eqref{app:total_kernel} at the points $\varphi$ and $\varphi'$ defining our discretization. These points act as a row and a column index, respectively. Similarly $\boldsymbol{f}$ and $\boldsymbol{j}$ are vectors with components $f(\varphi)$ and $j(\varphi)$, respectively (see Eq.~\eqref{eq:overlap} of the MT).
    
    The solution $\boldsymbol{f}$ is now used in a simple scalar product
    \begin{equation}
        D\ge\real\left(\frac{\boldsymbol{j'^\dagger}\cdot\boldsymbol{f}}{2}\right),
        \label{app:num2}
    \end{equation}
    which gives the lower bound (compare with Eq.~\eqref{eq:mazur} of the MT). Here $\boldsymbol{j'}$ is a permuted vector $\boldsymbol{j}$, its components being $j(\bar{\varphi})$ instead of $j(\varphi)$. Symbol $(\bullet)^{\boldsymbol{\dagger}}$ denotes the conjugate transpose.
    
    It appears that even relatively crude discretizations of roughly $L=200$ points along the relevant line provide excellent results, essentially indistinguishable from more precise discretizations. Still, we perform calculations with $L$ up to $5000$ to verify these results. 

    \vspace{\baselineskip}
    \textbf{Tensor network simulations.--}
    In order to check how well our bound describes the actual value
    we attempt to obtain numerical values for the spin Drude weight through an alternative approach. To this end, we simulate an inhomogeneous quench where the two halves of the system are initially prepared in a slightly polarized product  state
    \begin{equation}
        \rho_\mu=\frac{\left(e^{\mu_L\sigma^\z}\right)^{\otimes N/2}\otimes\left(e^{\mu_R\sigma^\z}\right)^{\otimes N/2}}{\tr{\left[\left(e^{\mu_L\sigma^\z}\right)^{\otimes N/2}\otimes\left(e^{\mu_R\sigma^\z}\right)^{\otimes N/2}\right]}}.
        \label{app:num3}
    \end{equation}
    We take $\mu_R=-\mu_L$ with $\mu_L=\mu/2$ typically being small (usually of order $10^{-2}$ or smaller), since this increases the stability of the simulation. 
    In order to obtain the Drude weight, we use the linear response expression
    \begin{equation}
        D=\lim_{t\to \infty}\lim_{N\to \infty}\lim_{\mu\to 0}\frac{\langle J(t)\rangle_\mu}{2 t \mu},
        \label{app:num4}
    \end{equation}
    recently used in \cite{vasseur15,karrasch17,ilievski17}, with $J$ being the total extensive current. We compute its expectation value using a TEBD algorithm. 
    Formula~\eqref{app:num4} differs from~\eqref{app:Kubo2}, derived in Appendix A, since this particular protocol gives rise to nontrivial dynamics only inside the light-cone, emerging from the contact between the two halves of the chain.
    
\begin{figure}[ht!]
        \centering
        \includegraphics[width=1.0\textwidth]{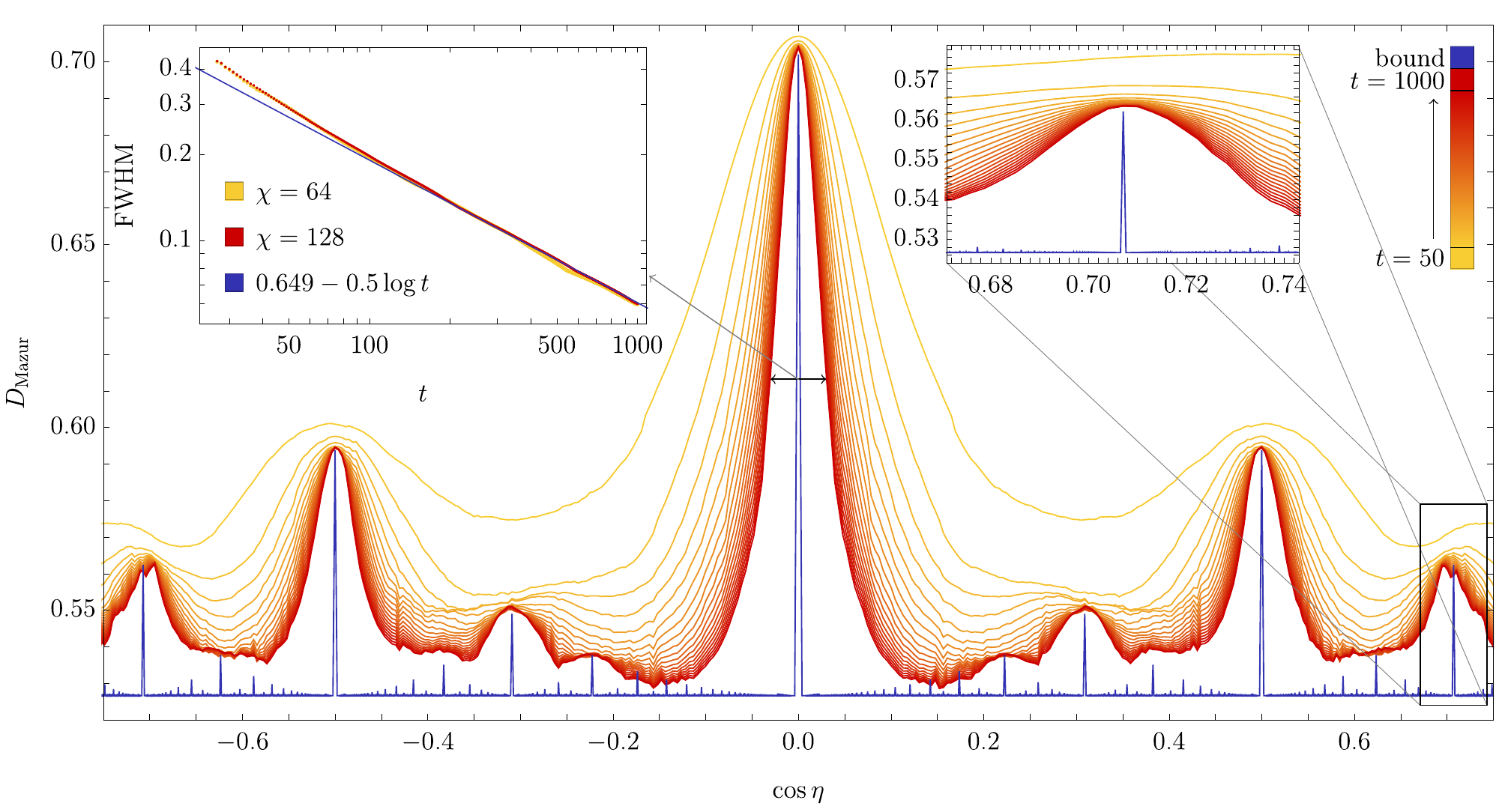}
        \caption{The unscaled fractal Drude weight at $|\lambda|=1$ as computed by two approaches: quasi-exact numerical solution of the Fredholm equation yielding the Mazur lower bound (blue) and the numerical TEBD/DMRG algorithm at long but finite times (yellow-red curves). The TEBD simulations in the main plot were performed using system size $N=3600$ and bond dimension $\chi=64$. The left inset shows the full width at half maximum (FWHM) of the TEBD results around the central peak (corresponding to commensurate anisotropy $\eta=\frac{\pi}{2}$). Comparing two different bond dimensions, it appears that FWHM converges to zero as $t^{-1/2}$ (left inset).
        The right inset shows more precise simulations (bond dimension $\chi=128$) for one of the regions where the TEBD results for $\chi=64$ seem to be below our strict theoretical lower bound. These occurrences are hence shown to be a result of small bond dimensions. Using larger bond dimension clearly improves the situation. } 
        \label{fig:app_num_01}
    \end{figure}

    Typically we simulate chains of length $N=3600$ and $N=7200$ if longer times are needed. The times we can reach are roughly a quarter of the chain's length, at which point the boundary effects may start affecting the accuracy of our Drude weight estimates. Finally, the size of the bias $\mu$ can be taken to be around $10^{-2}$. Taking a smaller magnetization step appears not to be necessary. For the most part, we take a bond dimension of $\chi=64$ which appears to be sufficient to obtain moderately accurate results with errors in the $\sim 1\%$ range (as compared to a somewhat more precise simulation at $\chi=128$ or even $\chi=256$ which were made occasionally for comparison). Increasing the bond dimension helps increase the Drude weight at points where the less precise calculation might duck under the theoretical lower bound. This can be seen at the peaks at commensurable $\eta/\pi$ ("the fractal spikes"), an example being shown in Figure \ref{fig:app_num_01}. 

    We stress that the numerical simulations clearly suggest convergence to a limiting fractal spin Drude weight as given by the Mazur bound. This convergence is quantitatively illustrated in the inset of Figure \ref{fig:app_num_01}, where the full width at half maximum (FWHM) of finite-$t$ DMRG data converges to zero as $\sim t^{-1/2}$.

\end{document}